\documentclass[12pt]{article}
\usepackage[dvips]{graphicx}
 \newcommand{\beq}{\begin{equation}}
 \newcommand{\eeq}{\end{equation}}
 \newcommand{\ber}{\begin{eqnarray}}
 \newcommand{\eer}{\end{eqnarray}}
\begin{document}
\title{QUANTUM WEAK TURBULENCE}

\author{Devashish Sanyal\thanks{tpds@mahendra.iacs.res.in} \\
Department of Theoretical Physics \\
Indian Association for the Cultivation of Science \\
Jadavpur, Calcutta 700032, INDIA \\
\and
Siddhartha Sen\thanks{sen@maths.tcd.ie}\\
School of Mathematics\\
 Trinity College \\
 Dublin 2 ,Ireland \\
 and \\
 Department of Theoretical Physics \\
 Indian Association for the Cultivation of Science \\
 Jadavpur, Calcutta 700032, INDIA}
\date{}
\maketitle
\begin{abstract}
 The study of the phenomenon of quantum weak turbulence is extended by determining the quasiparticle spectrum associated with such a system using a
 Green's function approach. The quasiparticle spectrum calculated
 establishes the dissipative regime and the inertial regime,
hence a Kolmogorov type of picture.

\vspace{.5cm}
\noindent
PACS number(s): 47.27.Ak, 67.40.Db, 67.40.Vs, 03.40.Gc

\end{abstract}

\section{Introduction}
As emphasized by Newell and Zakharov the phenomena of turbulence is
 not confined simply to fluid motion.
Weak turbulence in the general framework is the study of dispersive waves
(not necessarily in fluids) with weak non-linear
 interaction. Such a system can be represented  by a Hamiltonian
 corresponding to the wave equation with a small conservative non
linearity. The study of weak turbulence is relevant to many
 physical processes like capillary waves on the surface of water.

 It may be noted that the non-linear Schrodinger equation(NLS) $i\Psi_t + \triangle\Psi +\alpha
{\mid\Psi\mid}^s\Psi =0$, a candidate for weak turbulence, is widely used in continuum mechanics,
 plasma physics and optics. The NLS in the form $i\Psi_t + \triangle\Psi
+ \alpha{\mid\Psi\mid}^2 \Psi=0,\alpha=\pm 1$ describes a number of physical
 processes like waves on fluid surface etc. A consequence of weak turbulence is to allow a Kolmogorov like spectra (Kolmogorov-Zakharov spectra) which is exact[1,2]. In this paper
 the study of weak turbulence  in the quantum region described in Ref[3] is continued.
   With the help of a Green's function method the quasiparticle
 modes which describe excitations in the system are determined.  The
 quasiparticle spectrum establishes clearly the dissipative regime and
 hence the range of validity of the approach.
 \noindent
In the present study the starting Hamiltonian, in the language of second
 quantization, takes the form[3] 
\beq
H = {\sum_k \omega_k {a^\dagger}_k {a}_k} + {\sum_{k_1 k_2 k_3 k_4}T_{
k_1 k_2 k_3 k_4}{a^\dagger}_{k_1}{a^\dagger}_{k_2}a_{k_3}a_{k_4}}
\eeq
\noindent
  The  momentum conserving $d$ dimensional $\delta$ term has been absorbed in the
T matrix. The above Hamiltonian may be identified as the  microscopic
 Hamiltonian for a Bose liquid. The $
a^\dagger$ , $a$ are the usual creation and annihilation operators
 satisfying the  the commutation relation $[a_k ,{a^\dagger}_l ] =
 \delta_{kl}$. The mode number operator is given by $\hat{n_k}=
a^\dagger_k a_k$.
\section{Turbulence in a Bose Liquid}
 We interpret the state in which the occupation numbers are large as
 a fully developed turbulent system. The reason for this is that
 the system, as we will show, has a region ( inertial region)
 where there is no dissipation followed by a dissipative
 region at short distances.  In order to
 study turbulence we replace the $a_k$,${a\dagger}_k$ by
  c-numbers to the first approximation.This corresponds to assuming that the system is in a state in which there is large occupation number of different momenta.
 Such a system could model a driven system not in equilibrium relevant
 to turbulence. It may be emphasized here  that the liquid is not a 
 condensate which exists only at $k=0$. Such a condensate is postulated
  for  equilibrium calculations. 
 We divide the Hamiltonian into

   $ H= H_o + H_I$[3]. Since we are interested in the expectation values of
 fields between states with large mode numbers it is sensible to include
 as much  of these in the unperturbed Hamiltonian $H_o$ as the calculation
 permits.  We do  not need to assume that the coupling to all the quartic
 terms in the Hamiltonian is small; the diagonal part can be arbitrarily
 large in our approach. The unperturbed part($H_o$) thus  contains
    the quadratic part and the diagonal part  of the $T$ matrix i.e
     terms of the form $T_{klkl}$. After some simplification we can
      write
  \beq
  H_o = {\sum_k \omega_k \hat{n_k}} + 2{\sum_{k,l}
  T_{k,l}\hat{n_k}\hat{n_l}}
  \eeq
  \beq
  H_I = {\sum T'_{k_1 k_2 k_3 k_4}{a^\dagger}_{k_1}
  {a^\dagger}_{k_2}a_{k_3}a_{k_4}}
  \eeq
   where we have introduced the notation $T_k = T_{kkkk}$,$T_{kl} =
   T_{klkl}$ and
\beq
{T'_{k_1 k_2 k_3 k_4}} = \left\{ \begin{array}{r@{\quad:\quad}l}
{T_{k_1 k_2 k_3 k_4}} & k_1\neq k_3 or k_4 \\ 0 & otherwise
 \end{array} \right.
\eeq

\vskip0.5cm
 The next step is to calculate $\langle{\frac{d\hat{n_k}}{dt}}\rangle$
 , the quantum kinetic equation and look for the solution of the equation $\langle{\frac{d\hat{n_k}}
{dt}}\rangle = 0$( stationary solution). In order to establish Kolmogorov
  type solutions for the above equation the following scaling  relations
 have been assumed : $\epsilon (\lambda k) = \lambda ^{\alpha} \epsilon (k)$
,$T^{\prime}(\lambda k_1 ,\lambda k_2 , \lambda k_3 , \lambda k_4 ) =
 \lambda^{\beta} T^{\prime} ( k_1 , k_2 , k_3 , k_4 )$ and $N(\epsilon )
= \epsilon ^{z}$. $\epsilon_k $ is the energy corresponding to the mode
 k and $N(\epsilon)$ is the occupation number of mode having energy $\epsilon$.
 The solutions corresponding to the stationary
 state
 are given  by[3]
\beq
N^{(1)}(\epsilon) = C_1 \epsilon^{\frac{-(\triangle + 3)}{3}}
\eeq
\beq
N^{(2)}(\epsilon) = C_2 \epsilon^{\frac{-(\triangle + 4)}{3}}
\eeq
 where $\triangle = \frac{{3d + 2\beta}}{\alpha} - 4$.
\noindent
The solutions mentioned above do not correspond to the BE distribution,
 hence they are the non-equilibrium solutions.
We have ignored the two equilibrium solutions to the quantum
 kinetic equation which also exist. Having shown that such
 non-equilibrium scaling solutions exist we would like to analyse
 next the  behaviour of the unperturbed Hamiltonian
 for the  Kolmogorov type solutions as stated above. Note the unperturbed
 Hamiltonian in this approach contains the `` diagonal" part of the
 interaction as described in equation (2).
\section{ Energy Spectrum of the Unperturbed Hamiltonian $(H_o )$}
We are dealing with a Bose fluid,not in equilibrium, in which the occupation numbers of the  modes are very high,i.e $N_k\gg 1$. Hence
we now proceed to replace the operator ${a_k}$ by $ \sqrt {N_k} +
{b_k}$ and ${a^\dagger}_k$ by $\sqrt {N_k} +{b^\dagger }_k$where  $N_k$
 is a  c-number and the operators $b_k$,${b^\dagger}_k$ represent
 fluctuations around the ground state which have occupation $N_k$
 for the $k$th mode. This ground state will be represented by the 
 ket $\mid 0\rangle $.
 The above substitution results in the eqn ($3$) being rewritten  as
\beq
H_o = Const + ({\sum_k A_k {b^\dagger}_k b_k} + {\sum_k
B_k{b^\dagger}_k {b^\dagger}_{-k}} + {\sum_k B_k b_k b_{-k} })
\eeq
 where
\beq
 A_k = 4{\sum_l T_{kl} N_l}  + {\omega}_k + 4T_k N_k
\eeq
\beq
B_k = 4T_{k -k}N_k
\eeq
\noindent
 The approximate Hamiltonian,equation (7) constructed in this way can be called the Bogolyubov Hamiltonian. The energy spectrum for the above Hamiltonian may be found out  by
 using the method of  Green's functions as briefly described below.
 We define the Green's function ${G^{(1)}}_k = \langle 0 \mid T[{b^{\dagger
}_k}(t) b_k (0)]\mid 0\rangle$ and  ${G^{(2)}}_k = \langle 0 \mid T[b_
{-k}(t) b_k (0)]\mid 0\rangle$ where ${b^{\dagger}}_k (t)$, $b_{-k}(t)$
 are in Heisenberg representation. The   time ordered expectation
 values are taken between the  states in which there is macroscopic occupation number. This state is taken to represent the turbulent configuration and will be referred to as the ' ground state' of the system. The time derivatives of
${G^{(1)}}_k$ and ${G^{(2)}}_k$ yield,
\beq
\frac{d{G^{(1)}}_k}{dt} = -\delta (t) + 2i B_k {G^{(2)}}_k +
 i A_k{G^{(1)}}_k
\eeq
\beq
\frac{d{G^{(2)}}_k}{dt} = - 2i B_k {G^{(2)}}_k -i A_k {G^{(1)}}_k
\eeq
\noindent
 The fourier transform of the above equations  with respect to
 the variable t  and the evaluation of ${{\tilde{G}}^{(1)}}_k (\omega)$
 , the fourier transform of ${G^{(1)}}_k$ leads to
 \beq
  {{\tilde{G}}^{(1)}}_k (\omega) = -\frac{\omega + A_k}{\omega^2 -({A_k}^2 -
  4{B_k}^2 )}
  \eeq
  The poles of ${{\tilde{G}}^{(1)}}_k$ are given by $\omega =
  \pm {( {A_k}^2 -4{B_k}^2 )}^{1/2}$.  Hence the energy spectrum of the
   quasiparticles   is given by $ E = E_o + {\sum_k \epsilon (k)
    {c^\dagger}_k c_k}$ where
    \beq
    \epsilon (k) = \sqrt{(A_k - 2B_k)(A_k + 2B_k)}
    \eeq
    \noindent
    Here $E_o$ represents the ground state energy and $c_k$,${c^\dagger}_k$ are
 the annhilation and the creation operators for the quasi-particles. The
 equation $(14)$  may be solved self-consistently using the expressions
 $(9)$ and $(10)$ and the scaling form of the following functions:
      \noindent
      ${\omega}_{\lambda k} \sim \lambda^2 {\omega}_{k} $, $T_{\lambda k} \sim \lambda^\beta T_{k}$, $N_{\lambda k} \sim \lambda^{x} N_{k}$
    and $\epsilon (\lambda k) \sim \lambda^\alpha \epsilon (k)$. Let $T_{\lambda k} N_{\lambda k} \sim \lambda^{\gamma} T_{k} N_{k}$ where $\gamma =
 \beta + x $. If we  consider the term $A_k + 2B_k$, under recsaling of the
 momenta by $\lambda$, we have,
\noindent
$ A_{\lambda k} + 2 B_{\lambda k} = 4\lambda^{\gamma +1} \sum {T_{kl} N_{l}} +
4\lambda^\gamma T_k N_k + \lambda^{2} \omega_k$. In the infrared limit
$(k\rightarrow 0)$ and $\lambda\rightarrow 0$, it is seen that the
 term containing $\omega_k$ term is rendered ' irrelevant' for $\gamma<2$ and for $\gamma
>2$, $\omega_k$ is the only relevant term. Hence we have for asymptotically
small values of k
\beq
\epsilon (k) \sim \left\{ \begin{array}{r@{\quad:\quad}l}k^\gamma &
 \gamma < 2 \\ k^{2} & \gamma >2\end{array} \right.
\eeq
\noindent
Similar expressions can be written down for the large $k$ spectrum.
 Let us bring in  the concept of superfluidity here.
 Consider a body of mass M moving throught he fluid
 and in the proccess excites a quasiparticle of wave number
 $k$ and energy $\epsilon (k)$. Conservation of momentum
 yields,
\beq
M v_i = M v_f + hk
\eeq
For dissipation to occur the following relation amongst the energies should 
 be satisfied,
\beq
\frac{M {v_i}^2}{2} > \frac{M {v_f}^2}{2} + \epsilon (k)
\eeq
 The above two equations lead to the relation $v_i > \frac{\epsilon (k)}{k}$
 where $v_c = \frac{\epsilon (k)}{k}$  is called the critical velocity.
If  the velocity $v$ of  a particle moving through the ``fluid" of quasiparticles is less than $v_c$, where $v_c$ is the critical velocity
 , then the particle cannot loose energy,the fluid is termed to be a
 superfluid i.e the external particle moving through this
 fluid experiences no drag. This is the `` turbulent regime"
 for the system. In this regime there is no viscosity and hence
 corresponds to the large `` Reynolds number" situation.The order of magnitude of $v_c$ is given
 by $v_c\sim\frac{\epsilon (k)}{k} \sim k^{\alpha - 1}$. If $\alpha > 1 $ , it cannot be termed a
  superfluid as the critical velocity is zero. For $\alpha \leq 1$ it is a super
fluid as the critical velocity is very large for small $k$ values.Let 
 us consider an eddy of size $k$ which is associated with quasiparticle
 of wavenumber k. As is seen from the expression of $v_c$ the eddy
 critical velocity is $k$ dependent. The velocity scale of each eddy
 is set by a different $v_c$. The fluid analogy is further strengthened
 by the fact that dissipation occurs when $v > v_c $ i.e for increasing
 $k$. We will show that a solution where the  energy flow is towards
 the large k region exists. This is the turbulent  flow picture.

 Note this solution makes it clear that we are in a non-equilibrium
 situation. For an equilibrium configuration the system should evolve
 to it's lowest energy state(i.e small $k$ region). 
\\
Considering the total Hamiltonian H of equation (2), the resulting multiscale in  the  system would manifest in intermittency. 
 It is now evident from equation $(15)$   that
 $\alpha \le 1$ corresponds only to the $\gamma < 2 $ case from a self-consistent picture.We now
 equate $\alpha = \gamma $ to find out the $\alpha$ corresponding to the
 solution given by equations $(5)$ and $(6)$ . The solutions are,

\beq
\alpha = 3d - \beta + 3
\eeq
\beq
3d - \beta = -3
\eeq

\noindent
 As is evident from (18), the solution (6) is unphysical. It has been 
 explained later that the above fact is a consequence of the non-equilibrium
 nature of the problem.
 For superfluidity we have $\alpha \leq 1$. Imposing the condition on the
 equations $ (16)$  we obtain the lower bound on $\beta$ for
 the  K-Z solution   which is $\beta > 11$  for d=3. The
 result gives us the condition for superfluidity and also explicitly
  introduces a cut-off $(k_d)$ beyond which the fluid   ceases to
 be superfluid(turbulent) as the quasiparticle's k dependence follows
 from that of the interaction term $T_k$.
Since we are dealing with superfluid turbulence here  one may like to compare the
energy content in the eddy $k$ given by $E_k = \epsilon (k) N_k$ to the Kolmogorov's result in the fluid turbulence
, $E_k \sim k^{-{\frac{5}{3}}}$. It may be emphasized here that we are
 working in a region that is analogous to the  inertial regime in the
 fluid turbulence   and Kolmogorov's law is only valid in the
 inertial region.  The infra red region in our case is
 defined with respect to the  critical velocity of the system.

\vskip0.5cm

 For  consistency of the physical picture we need to study
 how the energy flows in the model following the techniques due to
 Zakharov[1]. The quantum
kinetic equation for the mode number
 per unit energy range, $N_{\epsilon}$, is given by[3],
\ber
\lefteqn{\langle{\frac{dN_\epsilon}{dt}}\rangle = } \nonumber\\
&& \int d\epsilon_1 d\epsilon_2 d\epsilon_3
  \big[U (\epsilon_1, \epsilon_2, \epsilon_3,\epsilon){(\epsilon_1 \epsilon_2
\epsilon_3 \epsilon)}^{-z}{ ( {\epsilon_3}^{z} + {\epsilon}^{z} -
{\epsilon_1}^{z} - {\epsilon_2}^{z})}{\epsilon^{-y}} \nonumber\\
&&  ( {\epsilon}^{y}
+ {\epsilon_3}^{y} - {\epsilon_2}^{y} - {\epsilon_1}^{y})
\delta(\epsilon + \epsilon_3 - \epsilon_2 -\epsilon_1 )\big]
\eer
 where $ y = 3z - p  -3 $,$N_{\epsilon}\sim \epsilon^{z}$ and $p = \frac{3d+2\beta}{\alpha} -4$.
\noindent
 Under the rescaling of the variables $\epsilon,\epsilon_1,\epsilon_2$
 and $\epsilon_3$   by $\lambda$ we have $U(\lambda\epsilon_1, \lambda
\epsilon_2,\lambda\epsilon_3,\lambda\epsilon) = \lambda^{p} U(\epsilon_1,
\epsilon_2,\epsilon_3,\epsilon) $ where $ p =  \frac{3d + 2\beta}{\alpha}
 - 4$. Denoting the above integral by $K(\epsilon , z,y)$ we have the
 following relation$
K(\epsilon,z,y) = \lambda^{-y -1} K(\lambda \epsilon,z ,y)$. Putting
$\lambda = \frac{1}{\epsilon}$. Hence we have
\beq
K(\epsilon ,z,y)= \epsilon^{ -y -1} K(1,z,y)
\eeq
We now define  the  current corresponding to the conservation equation
 for the mode number per unit energy range($N_{\epsilon}$) as $j$. The
 conservation equation is,
\beq
\langle{\frac{\partial N_\epsilon}{\partial t}}\rangle = \frac{\partial j}{\partial \epsilon}
 = \epsilon ^{-y -1} K(1,z,y)
\eeq
\beq
\Rightarrow  j = -\frac{\epsilon^{-y}}{y} K(1,z,y)
\eeq
  $j$ can be identified with the energy curent per unit energy range
 as can be seen from simple dimensional analysis. Considering the
 stationary solutions  i.e. K-Z solutions  we have for
 the case $ y=0$,
\beq
j = {-\frac{\partial K(1,v,y)}{\partial y}}
{\Big|_{y=0}} \mbox{ (using L'Hospital's rule)}
\eeq
 Considering the other stationary solution i.e y =1 we have,
\beq
j = 0
\eeq
 Since we are considering turbulence, a non-equilibrium phenomenon, only
 the solution corresponding to $j\not=0$ is the true solution while
 $j=0$ is the unphysical one. Thus the
 flux analysis rules out the second solution.The fact that $\alpha$ cannot
 be self-consistently determined from the second solution is manifest
 in the unphysical nature of the solution as discussed
 above.Note that the current corresponding to the non-zero solution
 of equation
 (21) represents an energy flux from low k end (wavnenumber at which energy is injected into the system) to the high k end where dissipation takes
 place in a manner very similar to one encounters in ordinary fluid
 turbulence governed by the Navier-Stokes equation {\it{(See Fig 1)}}.

 The results confirm that only one of the scaling solutions found  correspond to a  non-
equilibrium situation in which the energy flux is
non-zero.
\setlength{\unitlength}{1cm}
\begin{center}
\begin{picture}(8,8)
\put(0,4){\vector(1,0){6}}
\put(3,4){\line(0,1){.1}}
\put(1.5,4.5){\vector(1,0){1.4}}
\put(1.5,4.5){\vector(-1,0){1.4}}
\put(4.5,4.5){\vector(1,0){1.4}}
\put(4.5,4.5){\vector(-1,0){1.4}}
\put(0.5,4.8){\small{Superfluid}}
\put(3.4444,4.8){\small{Viscous}}
\put(0,3.5){\small$k=0$}
\put(5.5,3.5){\small$k\rightarrow\infty$}
\put(1.5,3){\vector(1,0){3.5}}
\put(1.3,2.6){\small{Direction of energy flow}}
\put(2.6,0.5){\small$Fig1$}
\end{picture}
\end{center}
 In a realistic situation there will be a driving source term for
 injecting energy into the system. Such systems have been extensively analysed
 by Zakharov[1]. It is clear that the arguments continue to hold for
 the general  quantum system we have considered. The analysis of the
 quasiparticle spectrum associated with quantum turbulence thus clearly establishes
 the inertial range, it places  restrictions on parameters,  demonstrates
 that scaling solutions which are not in equilibrium are possible and for the case $\alpha <1$ i.e where $v_c$ is k dependent suggests that a novel form of intermittency can occur. The analysis  thus completes the approach outlined in Ref[3]. We conclude with the following remarks. In the general approach of Zakharov to weak wave
 turbulence it is shown [1] that non-equilibrium scaling solutions are possible and that there is flux of either momentum or energy present in the
 system. Our  earlier analysis of the quantum version of Zakharov approach
 [3] established that these features are also present in the quantum
 system described. In this paper we have gone further. It has been shown
 that underlying such a non-equilibrium solution there is an inertial range with no dissipation
  and that a self consistently determined short distance dissipative scale can be
  found. Thus for the quantum system considered the analysis shows
 that the original intuition
 of the Kolmogorov approach to scaling can be justified.

\vskip0.5cm

\noindent
{\Large \bf Acknowledgements}
\vskip0.2cm
One of the authors(DS) would like to acknowledge the financial assistance
 given to him by  the Council of Scientific and Industrial Research, Gover
nment of India.

\end{document}